\documentclass{iau} 
\usepackage{graphicx}
\usepackage{hyperref}	
\hypersetup{pdfborder = {0 0 1}}
\usepackage{float}
\usepackage{graphicx}
\usepackage{times}
\usepackage{amstext}
\usepackage{amsmath}
\usepackage{amssymb}	
\usepackage{natbib}
\usepackage{url}
\usepackage{tabularx}
\usepackage{mathtools}
\usepackage{mathrsfs}
\usepackage{graphics}
\usepackage{multirow}
\usepackage{ar}
\usepackage{caption} 
\usepackage[rightcaption]{sidecap}
\usepackage{natbib}
\newcommand{\fdg}{\mbox{\ensuremath{.\!\!^\circ}}}%

\title[Great oaks from little acorns grow] 
{Molecular gas kinematics of the CMZ:\\ Great oaks from little acorns grow}

\author[Jonathan D. Henshaw]   
{Jonathan D. Henshaw$^{1}$}

\affiliation{$^1$Astrophysics Research Institute, Liverpool John Moores University, \\ Liverpool, L3 5RF, UK \\ email: {\tt j.d.henshaw@ljmu.ac.uk}}

\pubyear{2016}
\volume{322}  
\jname{The Multi-Messenger Astrophysics of the Galactic Centre}
\editors{R. Crocker, S. Longmore,  \& G. Bicknell }
\begin{document}

\maketitle

\begin{abstract}
The central molecular zone (CMZ) hosts some of the most massive and dense molecular clouds and star clusters in the Galaxy, offering an important window into star formation under extreme conditions. Star formation in this extreme environment may be closely linked to the 3-D distribution and orbital dynamics of the gas. Here I discuss how our new, accurate description of the $\{l,b,v\}$ structure of the CMZ is helping to constrain its 3-D geometry. I also present the discovery of a highly-regular, corrugated velocity field located just upstream from the dust ridge molecular clouds (which include G0.253+0.016 and Sgr B2). The extremes in this velocity field correlate with a series of massive ($\sim10^{4}$~M$_{\odot}$) cloud condensations. The corrugation wavelength ($\sim23$~pc) and cloud separation ($\sim8$~pc) closely agree with the predicted Toomre ($\sim17$~pc) and Jeans ($\sim6$~pc) lengths, respectively. I conclude that gravitational instabilities are driving the formation of molecular clouds within the Galactic Centre gas stream. Furthermore, I suggest that these seeds are the historical analogues of the dust ridge molecular clouds -- possible progenitors of some of the most massive and dense molecular clouds in the Galaxy. If our current best understanding for the 3-D geometry of this system is confirmed, these clouds may pinpoint the beginning of an evolutionary sequence that can be followed, in time, from cloud condensation to star formation.
\keywords{stars: formation -- ISM: kinematics and dynamics -- Galaxy: center }

\end{abstract}

\vspace{-0.5cm}

\section{Introduction}

The ultimate goal of star formation research is to develop an understanding of stellar mass assembly as a function of environment. The central molecular zone of the Milky Way (CMZ; i.e. the inner few hundred parsecs) hosts some of the most extreme products of the star formation process within the Galaxy. These include massive star clusters (e.g. Arches and Quintuplet; \citealp{longmore_2014}), massive ($10^{5}-10^{6}$\,M$_{\odot}$) protocluster clouds (e.g. Sgr B2), and massive ($10^{4}-10^{5}$\,M$_{\odot}$) infrared dark clouds (e.g. G0.253+0.016; \citealp{longmore_2012,rathborne_2014,rathborne_2015}). In spite of this, the present-day star formation rate ($\sim0.05$\,M$_{\odot}$\,yr$^{-1}$; e.g. \citealp{crocker_2012}) is 1-2 orders of magnitude lower than one might naively expect when only considering the reservoir of dense ($\gtrsim10^{3}$\,cm$^{-3}$) gas (\citealp{longmore_2013a, kruijssen_2014b}).

Star formation within this complex environment may be closely linked to the orbital dynamics of the gas (\citealp{longmore_2013b}). Understanding how this influences star formation necessitates an accurate, holistic understanding of the gas kinematics and distribution -- something which has been sought after for several decades (e.g. \citealp{bally_1988, sofue_1995, jones_2012}). To date, the kinematics of the CMZ have been described using a combination of position-velocity diagrams, channel maps, and moment analysis. Although such techniques are simple to implement and well-understood, their output can be subjective and misinterpreted in complex environments.

With this in mind, \citet[H16]{henshaw_2016a} developed an analysis tool which efficiently and systematically fits large quantities of multi-featured spectral line profiles with multiple Gaussian components. Using {\sc scouse}\footnote{Publicly available at \url{https://github.com/jdhenshaw/SCOUSE}} H16 were able to investigate the kinematics of the CMZ on the scales of individual clouds, \emph{whilst maintaining a global perspective} -- critical in understanding how environment influences star formation in this complex region.


\sidecaptionvpos{figure}{c}
\begin{SCfigure}
\captionsetup{format=plain}
\includegraphics[trim = 10mm 5mm 10mm 10mm, clip, width = 0.40\textwidth]{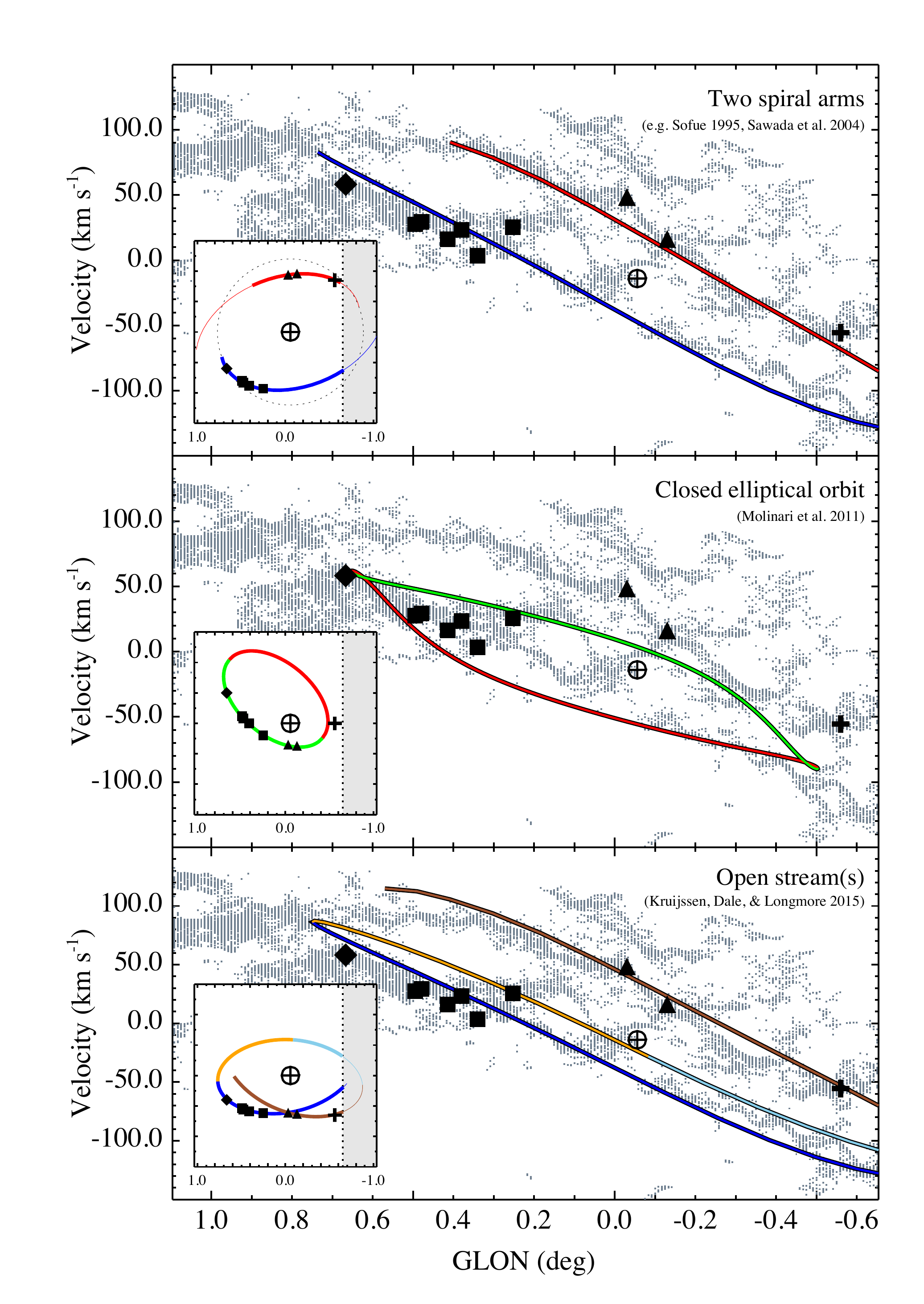}
\vspace{-0.23cm}
\caption{\small{Three different interpretations for the 3-D structure of the CMZ as they would appear in $\{l,\,v_{\rm LSR}\}$ space. Within each panel the inset figure represents a schematic of the top-down view of the respective interpretation. The top, central, and bottom panels refer to the spiral arm, closed elliptical, and open stream interpretations (see text), respectively. The black circle with a plus denotes the location of Sgr A*. Additionally, the locations of prominent molecular clouds are overlaid. In order of increasing Galactic longitude (from right to left): The Sgr C complex (black plus); the 20~km\,s$^{-1}$ and 50~km\,s$^{-1}$ clouds (black upward triangles); G0.256+0.016 (black square); Clouds B-F (black squares); The Sgr B2 complex (black diamond). }}
\label{Figure:pv_models}
\end{SCfigure}

\vspace{-0.5cm}

\section{Constraining the 3-D geometry of the CMZ gas stream}

Each panel in Fig.~\ref{Figure:pv_models} depicts a different interpretation for the 3-D geometry of the CMZ, as it appears in $\{l,\,v_{\rm LSR}\}$-space, in comparison to the {\sc scouse}-derived kinematics (H16). The inset image in each panel presents a face-on schematic of each interpretation (rotation is clockwise and direction to Earth is towards the bottom of the page). 

\vspace{0.1cm}

\noindent{\textbf{Spiral arms (top-panel):}} It has been suggested that the gas is distributed throughout two prominent spiral arms (e.g. \citealp{sofue_1995, sawada_2004}). The blue and red lines depict gas on the near- and far-side of the Galactic Centre (black circle with cross), respectively. In this interpretation, the dust ridge molecular clouds (from the square, G0.253+0.016 to the diamond, Sgr B2) are associated with the near-side arm, whereas Sgr C molecular cloud complex (plus symbol) and the 20~km\,s$^{-1}$ and 50~km\,s$^{-1}$ clouds (triangles) are linked to the far-side arm (see inset image).

\vspace{0.1cm}

\noindent{\textbf{Closed-elliptical orbit (central-panel):}} The specific parameterisation of a closed elliptical orbit shown in Fig.~\ref{Figure:pv_models} was made by \citet[M11]{molinari_2011}. In contrast to the spiral arm view, the 20~km\,s$^{-1}$ and 50~km\,s$^{-1}$ clouds are situated on the near-side of the Galactic Centre. Moreover, the gas associated with these clouds connects continuously to the dust ridge molecular clouds.

\vspace{0.1cm}

\noindent{\textbf{Open stream (bottom-panel):}} In a more recent view, developed via dynamical modelling of the system by \citet[KDL15]{kruijssen_2015}, the gas follows an open, rather than a closed orbit. Again, focussing on the 20~km\,s$^{-1}$ and 50~km\,s$^{-1}$ clouds, these are situated on the near-side of the Galactic Centre (as in the closed-elliptical configuration), but there is no immediate connection to the dust ridge, contrary to the closed-orbit geometry of M11.

\vspace{0.1cm}

The spiral arm interpretation provides a qualitatively good match to the kinematics. However, quantitative comparison cannot be made as there is currently no physical model of this system. The geometry was initially inferred empirically \citep{sofue_1995}, and later revisited by \citet{sawada_2004}, who used a combination of emission and absorption lines to determine the placement of clouds along the line-of-sight (LOS). This was attempted using coarse angular-resolution data ($0\fdg2\sim30$\,pc), and placement of the 20~km\,s$^{-1}$ and 50~km\,s$^{-1}$ clouds on the far-side of the Galactic Centre is inconsistent with both clouds appearing as absorption features at 70~$\mu$m (M11). Agreement would require no physical connection between these clouds and the far-side arm, which seems unlikely given the large-scale coherency and continuity of the $\{l,\,v_{\rm LSR}\}$ structure. The contrasting placement of these clouds on the near-side of the Galactic Centre in the M11 interpretation, and the inferred connection to the dust ridge molecular clouds, however, leads to a poor reproduction of the gas kinematics. The continuity and coherency of the two dominant, almost parallel, $\{l,\,v_{\rm LSR}\}$ features (Fig.~\ref{Figure:pv_models}), renders connection between these two groups of clouds physically problematic. Finally, the open stream interpretation provides a good match to the data, with the kinematics having been used as a fundamental diagnostic in a parameter space study of possible orbits in the known gravitational potential (KDL15).

Work is currently underway to obtain proper motions of masers with the VLBA (Immer~et~al. in~preparation). CH$_{3}$OH and H$_{2}$O masers, best suited for star-forming regions, have been detected in Sgr B2, Sgr C, the 20~km\,s$^{-1}$ cloud, south of G0.253+0.016, and dust ridge clouds `c' and `e' (e.g. \citealp{caswell_2010, walsh_2014,lu_2015}). Differences in the presumed LOS placement of molecular clouds, leads to observationally testable predictions of their motion. For instance, placement of the 20~km\,s$^{-1}$ cloud on the far-side of the Galactic Centre in the spiral arm interpretation, leads to a predicted plane-of-the-sky motion directed towards negative Galactic longitudes. The opposite is true for the other two interpretations. Results distinguishing between the different models can be expected within the next few years.


\vspace{-0.4cm}

\section{Seeding the gas stream: the formation of protocluster clouds}

Observations of the dust ridge clouds, which show an increase in star formation activity as a function of increasing Galactic longitude, led \citet{longmore_2013b} to suggest that these clouds may share a common timeline. The proposed evolutionary sequence begins with G0.253+0.016, which displays very little star formation activity (e.g. \citealp{mills_2015}), through to Sgr B2, which displays prominent star formation activity. In this scenario, star formation is initiated at a common point, triggered as the gas clouds are tidally compressed during their pericentre passage at $\sim$60~pc from the bottom of the Galactic gravitational potential (KDL15).

\begin{SCfigure}
\captionsetup{format=plain}
\includegraphics[trim = 0mm 0mm 0mm 30mm, clip, width = 0.39\textwidth]{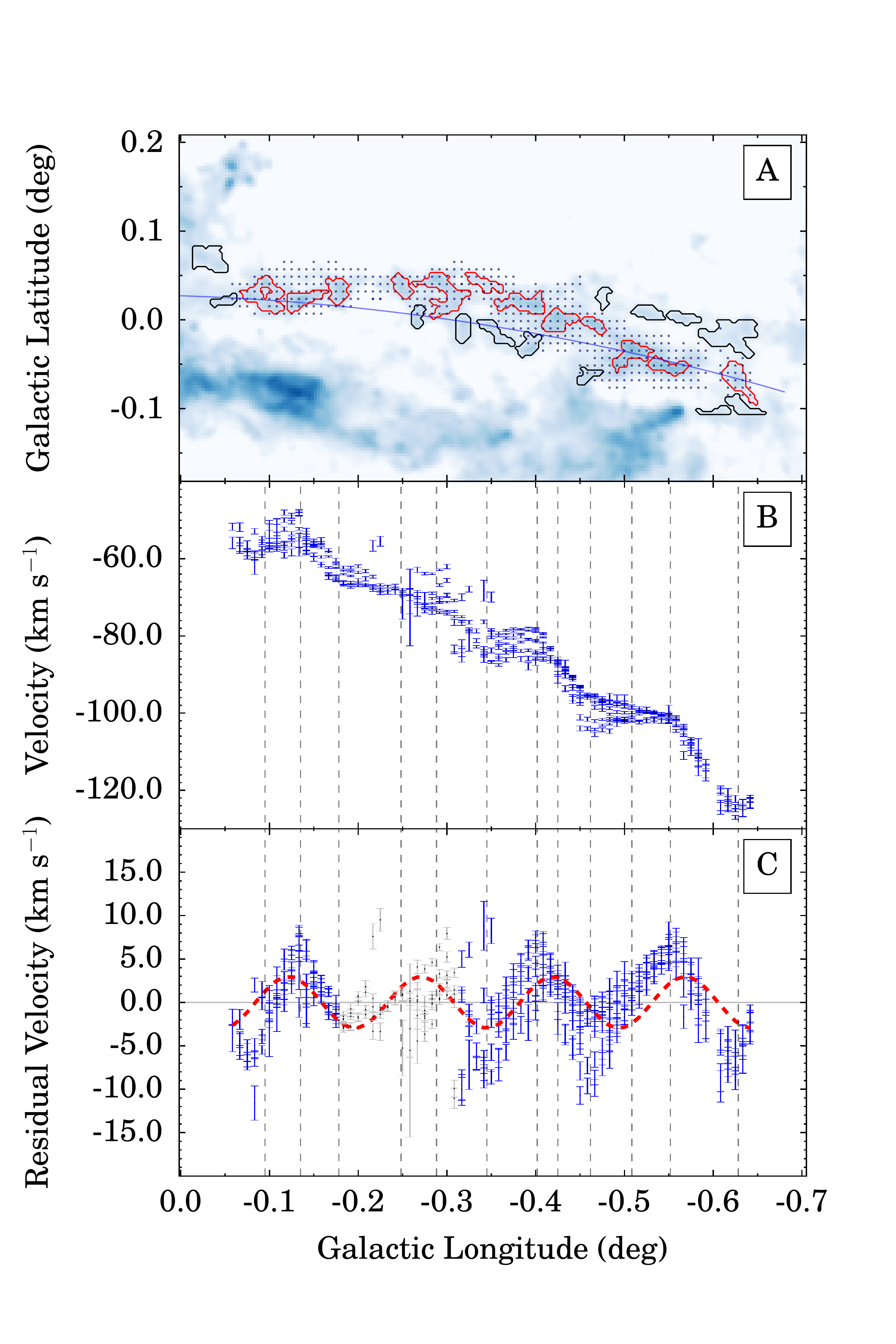}
\vspace{-0.5cm}
\caption{\small{Galactic longitude versus: (A) Galactic latitude. A column density map of the CMZ focusing on the region of interest (Battersby et al. in preparation). The blue solid line is equivalent to that in Fig.~\ref{Figure:pv_models}. The contours refer to molecular cloud condensations. The red contours are those spatially associated with the corrugated velocity field seen in panel B; (B) $v_{\rm LSR}$. The data points refer to the centroid velocities of spectral components extracted using {\sc scouse} \ (H16; blue data points in panel A); (C) residual velocity after removing a linear velocity gradient. The red dashed line represents the best-fitting solution to a sinusoidal model used to describe the blue data points. Grey data points indicate those that are excluded from the fitting process.} }
\label{Figure:wiggles}
\end{SCfigure}

An interesting question is whether or not it is reasonable to assume that the clouds have similar initial conditions, such that they evolve and form stars on similar time-scales once collapse has been initiated. To answer this question, \citet[HLK16]{henshaw_2016b} investigated the gas situated upstream from pericentre, identifying a highly-regular corrugated velocity field in $\{l, v_{LSR}\}$-space (Fig.~\ref{Figure:wiggles}). After correcting for inclination and orbital curvature (KDL15; see above and panel A), HLK16 derive the amplitude and wavelength of the corrugation, finding $A = 3.7\pm0.1$~km\,s$^{-1}$ and $\lambda_{\rm vel,i} = 22.5\pm0.1$~pc, respectively (red dashed line; Fig.~\ref{Figure:wiggles}). 

The correspondence between the extremes in the corrugated velocity field and a series of quasi-regularly spaced ($\lambda_{\rm sep,i}\sim8$~pc) cloud condensations (panel A), with typical masses $\sim10^{4}$~M$_{\odot}$, implies that gas is accumulating where the local velocity differential is minimal. HLK16 investigate the hypothesis that these `traffic jams' may lead to the development of gravitational instabilities. On size scales smaller than the Jeans length gas is stabilized against gravitational collapse by internal pressure, and on those greater than the Toomre length the gas is stabilized by rotation and shear \citep{toomre_1964}. Measuring the velocity dispersion, $\sigma_{v}\sim~5$~km\,s$^{-1}$, and estimating the gas surface density, $\Sigma\sim1000$~M$_{\odot}$pc$^{-2}$, in this portion of the stream, as well as the epicyclic frequency, $\kappa\sim3.2$~Myr$^{-1}$ \citep{launhardt_2002}, the turbulent Jeans and Toomre lengths are $\lambda_{\rm J}\approx\sigma_{v}^{2}/G\Sigma\sim6$~pc and $\lambda_{\rm T}\approx4\pi^{2}G\Sigma/\kappa^{2}\sim17$~pc, respectively. Since the smallest gas condensations are separated by a Jeans length ($\lambda_{\rm sep,i}\sim\lambda_{\rm J}$) and the corrugation scale matches the Toomre length ($\lambda_{\rm vel,i}\sim\lambda_{\rm T}$), HLK16 confirm that structure is growing within the stream on the same scales gravitational instabilities are expected to develop. Gravitational instabilities are therefore likely driving the formation of structure within the gas stream. 

Observations of the dust ridge molecular clouds reveal these clouds are similarly spaced ($\sim~12$~pc), and the surface density contrast against the stream is a factor of $\sim5$ greater in the dust ridge. Coupling this with the dearth of star formation activity reported towards the observed condensations, suggests these newly identified clouds are in an early evolutionary phase. Placing this information back into the global context, it is worth reaffirming that these two groups of clouds appear to be part of a contiguous gas stream that extends over 150~pc in projection (H16). In the KDL15 model, the condensations are 0.3--0.8 Myr upstream from G0.253+0.016, and hence may eventually reach the masses and densities of the dust ridge clouds by accreting more gas over the next free-fall time ($\langle t_{\rm ff}\rangle\sim0.5$\,Myr). In this scenario, the condensations may be historical analogues to the dust ridge clouds -- and therefore possible progenitors of some of the most massive and dense molecular clouds in the Galaxy. If confirmed, these clouds may pinpoint the initial phases of star and cluster formation; the beginning of an evolutionary sequence that we can follow, as a function of time, from cloud condensation to star formation.

\sidecaptionvpos{figure}{c}


\vspace{-0.5cm}


\bibliographystyle{IAU_JDH}
\bibliography{references} 

\end{document}